\documentclass[12pt]{article}
\usepackage{amsmath,amssymb}
 \usepackage{graphicx}
 
\begin{document}

 \begin{center}

{\large\bf  Surface response of spherical core-shell structured nanoparticle
by optically induced elastic oscillations of soft shell
against hard core}

\vspace*{.50cm}

S.I. Bastrukov\footnote{Corresponding author,
 e-mail: bast@jinr.ru}$^3$, P-Y. Lai$^{2}$, I.V. Molodtsova$^1$, H-K. Chang$^3$ D.V. Podgainy$^1$

\vspace*{0.5cm}
 $^1$Laboratory of Informational Technologies, Joint Institute for Nuclear Research,
 141980 Dubna, Russia

 $^2$ Department of Physics and  Center for Complex Systems, National Central University,
 Chungli 320, Taiwan

$^3$ Department of Physics and  Institute of Astronomy, National Tsing Hua University,
 Hsinchu 30013, Taiwan

 \end{center}

\vspace*{1.0cm}

\begin{abstract}
  The optically induced oscillatory response of a spherical two-component, shell-core structured, nanoparticle by nodeless elastic vibrations of soft peripheral shell against hard and dynamically immobile inner core is considered.
  The eigenfrequencies of the even-parity, spheroidal and odd-parity torsional vibrational modes trapped in the finite-depth shell are obtained which are of practical interest for modal
  specification of individual resonances in spectra of resonant scattering of long wavelength ultrafine particles.
\end{abstract}

\vspace{2.cm}

\section{Introduction.}
 Current increasing interest in the peculiarities of electromagnetic response
 of ultrafine particles (nano and micro dimensions) of different materials is motivated by their practical
 utilization for biomedical purposes. The well-know example is the nanoparticles of noble metals, like
 gold and silver, whose capability of responding by surface plasmon resonances (optically induced oscillations of
 valence electrons against immobile ions well-recognizable in the optical spectra of resonant photoabsorption)
 is of practical interest in the usage of these metallic nanoparticles as biolabels (e.g [1-3]).
 In \cite{SRL-06,PLA-05}, the model of electromagnetic response of a metallic nanoparticles placed in a permanent
 magnetic field  by surface gyromagnetic plasmons (optically induced cyclotron oscillations of
 valence electrons about the lines of magnetic field threading the nanoparticle) have been theoretically studied
 with concomitant suggestion that the considered effect can too be utilized in biomedical by applications.
 Continuing investigations in this direction, in present paper we consider two-component, hard core - soft shell, model of dielectric nanoparticle responding to externally induced electromagnetic load by shear elastic oscillations of the peripheral soft shell against hard dynamically inert core. The focus is laid on some technical aspect of solid-mechanical calculations of frequency of resonant oscillations. Specifically, our prime purpose is to elucidate how the frequencies of optically induced elastic shell-against-core oscillations (which in resonant
 photo absorbtion equal the frequency of impinging on particle AC electromagnetic field) depend upon
 material parameters and size of nanoparticle. Before embarking into details, it may be
 worth noting that theoretical treatment of experiments on electromagnetic response of ultrafine particles with relaxed surface from point of view of solid-mechanical model of elastic oscillations of soft finite-depth surface
 layer against hard core has first been considered, to the best of our knowledge, in work \cite{TI-83}.
 In this paper we consider non-studied before regime of nodeless shear elastic shell-against-core oscillations
 which is interesting in its own right and, first of all, because the obtained for the this case analytical
 expressions for frequencies of two fundamental vibrational modes, spheroidal and torsional, oscillations
 of shell against core clearly exhibit the set of characteristic parameters of nanoparticle.

\section{General equations of elastic shell-against-core nodeless vibrations}

 In what follows we study Rayleigh's regime of electromagnetic response of heterogenic shell-core structured
 spherical solid particle of radius $R$ to long wavelength electromagnetic field by non-compressional, pure shear, elastic oscillations of surface shell of finite depth $\Delta R=R-R_c$ against inner static core of radius $R_c$.
 This means that the wavelength of electromagnetic field is much large than spacing between sites of crystalline
 structure of particle (and may be even larger than the particle sizes) so that the particle material
 can be regarded in approximation of incompressible continuous medium.
 The fact that oscillations of peripheral shell are not accompanied by fluctuations
 in density means $\delta\rho=-\rho\,\nabla_k\,u_k=0$, where $u_i({\bf r},t)$ is the material displacements in the
 peripheral shell. This is the case  when
 crystalline structure undergo solely shear (reversal) distortions obeying the Hooke's law
 of linear relation between tensors of shear stresses $\sigma_{ik}$ and shear
 strains $u_{ik}$. All the above means that optically induced oscillations can be modeled by the standard
 equation of solid-mechanics for elastically deformable continuous medium
 \begin{eqnarray}
 \label{e2.1}
&& \rho{\ddot u}_i=\nabla_k\sigma_{ik}\quad \sigma_{ik}=2\,\mu\,
u_{ik}\quad\quad u_{ik}=\frac{1}{2}[\nabla_i u_k+\nabla_k u_i].
 \end{eqnarray}
The density $\rho$ and the shear modulus $\mu$ of particle material are regarded as
constant input parameters of the model.
The energy of oscillations is controlled by
equation
\begin{eqnarray}
 \label{e2.2}
  \frac{\partial }{\partial t}\int \frac{\rho {\dot u}^2}{2}\,d{\cal
  V} = -\int \sigma _{ik}{\dot u}_{ik}\,d{\cal V}=-2\int \mu\, u_{ik}{\dot u}_{ik}d{\cal
  V}.
   \end{eqnarray}
 As was said in introduction we focus on the regime of nodeless shear elastic oscillations
 in which fluctuations of material displacements ${\bf u}({\bf r},t)$ are described by solutions of
 the vector Laplace equation
 \begin{eqnarray}
 \label{e2.3}
 \nabla^2 {\bf u}({\bf r},t)=0\quad\quad \nabla\cdot{\bf u}({\bf
 r},t)=0.
 \end{eqnarray}
 In order to compute frequencies of nodeless oscillations in two fundamental modes - spheroidal and torsional,
 the material displacement ${\bf u}({\bf r},t)$ can be conveniently represented in the form
\begin{eqnarray}
 \label{e2.4}
 {\bf u}({\bf r},t)={\bf a}({\bf r})\,{\dot \alpha}(t).
 \end{eqnarray}
On substituting (\ref{e2.4}) in (\ref{e2.2}) we arrive at equation
for ${\alpha}$ having the form of standard equation of linear
oscillations
\begin{eqnarray}
 \label{e2.5}
 && \frac{dE}{dt}=0\quad E=\frac{M{\dot\alpha}^2}{2}+\frac{K{\alpha}^2}{2}
 \quad\to\quad {\ddot\alpha}+\omega^2\alpha=0\quad\quad \omega^2=\frac{K}{M}\\
 \label{e2.6}
 && M=\int \rho\, a_i\,a_i\,d{\cal V}\quad\quad
 K=2\int \mu\, a_{ik}\,a_{ik}\,d{\cal V}\quad  \quad a_{ik}=\frac{1}{2}[\nabla_i a_k + \nabla_k
 a_i]
 \end{eqnarray}
where the above introduced the field of instantaneous displacement
$a_i$ obeys too the Laplace equation
\begin{eqnarray}
 \label{e2.7}
 \nabla^2 a_k({\bf r})=0\quad\quad \nabla_k\, a_k({\bf r})=0
 \end{eqnarray}
which follows from substitution of (\ref{e2.4}) in (\ref{e2.3}). Two
fundamental vibrational modes in an elastic spherical shell,
spheroidal and torsional, are described by two fundamental solutions
of (\ref{e2.3}) built on general solution to the scalar Laplace
equation
\begin{eqnarray}
 \label{e2.7}
 \nabla^2 \chi({\bf r})=0\quad\quad
 \chi({\bf r})=f_\ell({\bf r})P_\ell(\zeta)\quad\quad
 f_\ell({\bf r})=[{\cal A}_\ell\,r^\ell+{\cal
  B}_\ell\,r^{-(\ell+1)}]
 \end{eqnarray}
 In the coordinate system with polar axis, the fields of displacements in the spheroidal mode ${\bf a}_s$ and
 torsional mode ${\bf a}_t$ of shear elastic vibrations are described by the
 poloidal and toroidal fields
 \begin{eqnarray}
 \label{e2.8}
&& {\bf a}_s=\nabla \times \nabla\times ({\bf
 r}\,\chi)\quad\quad {\bf a}_t=\nabla \times ({\bf r}\chi)\\
&&
 \chi({\bf r})=f_\ell({\bf r})P_\ell(\zeta)\quad\quad
 f_\ell({\bf r})=[A_\ell\,r^\ell+
  B_\ell\,r^{-(\ell+1)}]
 \end{eqnarray}
 respectively.
 Henceforth $P_\ell(\cos\theta)$ stands for the Legendre polynomial of multipole
 degree $\ell$ and arbitrary constants  $A_{\ell}$ and $B_{\ell}$
 must be eliminated from boundary conditions of the core-shell interface and on the particle surface.
 Remarkably, these above fields ${\bf a}_s$ and ${\bf a}_t$ as function of radial coordinate $r$
 have no nodes along the shell thikness $R_c < r <R$. The corresponding vibrational modes are specified, thereby,
 as modes of nodeless irrotational spheroidal oscillations (it can easily be verified that $\nabla\times {\bf a}_s=0$)
 and differentially rotational torsional oscillations. The
 case of nodeless global oscillations, in entire volume of spherical mass of a viscoelastic
 solid, pictured in Fig. 1, has been considered in details in [6].

\begin{figure}
\centering{\includegraphics[width=12cm]{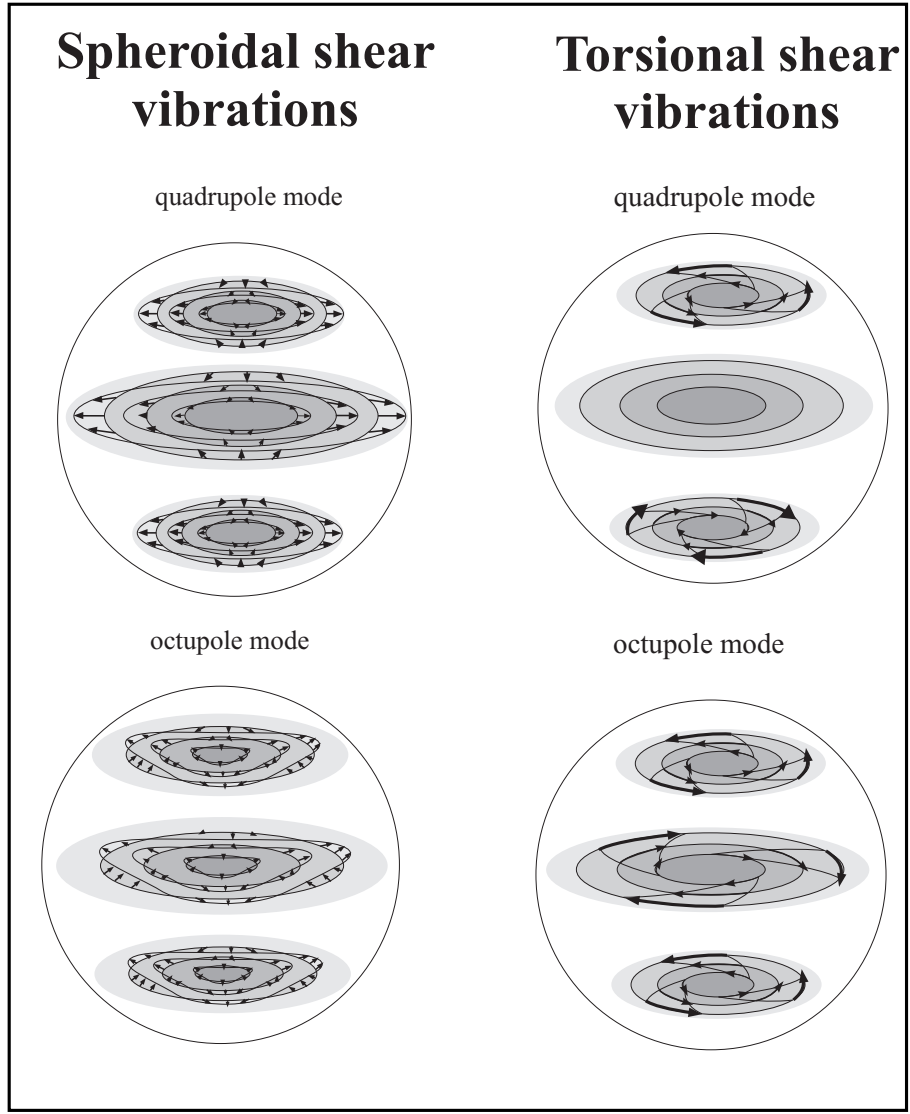}} \caption{
 Material displacements (arrows) in spherical mass of
elastic matter  undergoing nodeless spheroidal and torsional
vibrations of quadrupole $\ell=2$ and octupole $\ell=3$ degree. In
spheroidal mode with frequency
$\omega_s(\ell)=\omega_0[2(2\ell+1)(\ell-1)]^{1/2}$ and in torsional
mode with frequency
$\omega_t(\ell)=\omega_0[(2\ell+3)(\ell-1)]^{1/2}$, where natural
unit of frequency $\omega_0=c_t/R$ with $c_t=[\mu/\rho]^{1/2}$ being
the speed of transverse wave of elastic shear in the bulk
elastically deformable material of particle and $R$ is the particle
radius.}
\end{figure}

 In this work we derive spectral formulae for both
 spheroidal and torsional vibrational modes trapped in the
 peripheral  shell. Following the above outlined computational scheme we compute
 separately the frequencies of elastic spheroidal and torsional shear oscillations of the
 soft shell against hard core.

 \section{Nodeless spheroidal oscillations of soft shell against hard core}

 In case spheroidal response the arbitrary constants $A_\ell$ and $B_\ell$ can be eliminated
 from condition of impenetrability of perturbation in the core, $u_r\vert_{r=R_c}=0$,
 and condition of compatibility of rate of radial component of poloidal displacements
 with the rate of harmonic spheroidal distortions of the particle
 surface, ${\dot u}_r\vert_{r=R_c}={\dot R}(t)$, where $R(t)=R[1+\alpha(t)\,P_\ell(\cos\theta)]$.
 The computed along the above scheme frequency of spheroidal nodeless oscillations is given by
\begin{eqnarray}
\label{a1.12}
&&\omega^2_s(\ell,\lambda)={\omega_0^2}\,\frac{2(2\ell+1)}{(1-\lambda^{2\ell+1})}
\left[\frac{(\ell^2-1)(1-\lambda^{2\ell-1})+\ell(\ell+2)\lambda^{2\ell-1}(1-\lambda^{2\ell+3})}
 {(\ell+1)+\ell\lambda^{2\lambda+1}}\right]\\ \nonumber
 \\
 \label{e2.12a}
 && \omega_0^2=\frac{c_t^2}{R^2}=\frac{\mu}{\rho}\quad \lambda=\frac{R_c}{R}=1-h\quad h=\frac{\Delta R}{R}.
\end{eqnarray}
where $c_t=[\mu/\rho]^{1/2}$ is the speed of transverse wave of
elastic shear in the bulk of the shell material.
Note, that geometrical parameter $\lambda$ strongly less than unit,
$\lambda < 1$. Deserved for particular comment is
the dipole overtone of these oscillations
which possesses properties of Goldstone's soft mode. To see this, consider
the limit of zero-size radius of the core,
$\lambda=(R_c/R)\to 0$, which corresponds
global nodeless spheroidal elastic shear vibrations in entire spherical volume of particle.
In this limit we arrive at known result (e.g.[6-8])
\begin{eqnarray}
 \label{e2.13}
  \omega^2_s((\ell\geq 2,\lambda=0)=\omega^2_0[2(2\ell+1)(\ell-1)]
 \end{eqnarray}
which shows that lowest overtone of global, in entire volume, of
nodeless spheroidal oscillations is of quadrupole degree, $\ell=2$.
However, this is not the case when we are oscillations trapped in the peripheral layer of
finite depth. In this latter case from equation (\ref{e2.12})
it follows that lowest overtone is of dipole degree and the frequency of this dipole vibration
is given by
\begin{eqnarray}
  \label{e2.14}
  \omega^2_s(\ell=1,\lambda)=\omega_0^2\frac{9\lambda(1-\lambda^5)}{(1-\lambda^3)(1+\lambda^3/2)}.
 \end{eqnarray}
The dipole spheroidal elastic oscillations of soft layer
against inert hard are pictured in Fig. 2.

\begin{figure}
\centering{\includegraphics[width=10cm]{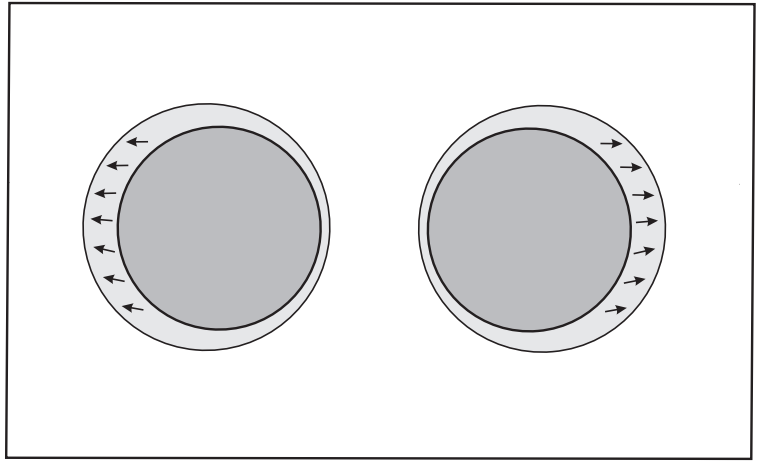}} \caption{
 Irrotational displacements in dipole spheroidal soft mode of shear elastic oscillations of peripheral shell
 against core.
}
\end{figure}

Thus, in case of global oscillations, in the whole volume, the
dipole overtone of spheroidal mode disappears what is means that the
frequency of dipole overtone tends to zero, as $\lambda\to 0$. This
behavior of fractional frequency of dipole spheroidal irrotational
oscillations of peripheral shell against core as a function of
fractional thickness $h=\Delta R/R$  is shown in Fig. 3.

\begin{figure}
\centering{\includegraphics[width=8cm]{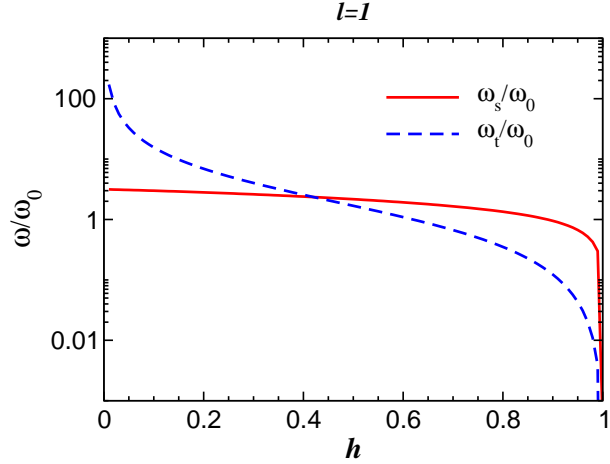}} \caption{
Fractional frequencies of dipole, $\ell=1$, spheroidal
  $\omega_s/\omega_0$ and  torsional $\omega_t/\omega_0$ oscillations
  of surface shell against hard core as functions of its fractional thickness $h=(R-R_c)/R$.
  When $h\to 1$, the radius of core $R_c\to 0$. The smooth decreasing of frequency
  shows the more mass (volume) of nanoparticles sets in dipole vibrations of shell against
  core the less frequency of these vibrations. The Goldstone's feature of this soft
  mode is that when all the mass of elastic matter
  of particle involves in oscillations, the dipole oscillatory excitation does not appear.}
\end{figure}

In this sense the dipole vibration can be specified regarded as, so
called, Goldstone soft mode, whose most conspicuous feature is to
vanish, when some parameter of oscillating system turn to zero. In
the model under consideration this parameter is $\lambda=(R_c/R)$.
Thus, the emergence of dipole overtone in the long wavelength
oscillatory response of two-ply particle is the most striking
feature of its vibratory behavior as compared to homogeneous
particle.

\section{Nodeless torsional oscillations of soft shell against hard core}

 In torsional mode of differentially rotational oscillations of elastically deformable soft shell against hard inert
 core, the constants $A_\ell$ and $B_\ell$ entering in expression for the field of toroidal instantaneous
 displacements, ${\bf a}_t$, are eliminated from
 the no-slip condition on the core-shell interface, $u_\phi\vert_{r=R_c}=0$,
 and on the globe surface $u_{\phi}\vert_{r=R}=[\mbox{\boldmath $\Phi$}\times {\bf R}]_\phi$,
 where $\mbox{\boldmath $\Phi$}={\alpha}(t)\nabla_{r=R} P_\ell(\zeta)$; the last condition
 is dictated by the form of general toroidal field whose correctness
 is discussed below. The resultant expression for the frequency
 of torsional oscillations of the soft layer against hard core is given by
 \begin{eqnarray}
\label{e2.15}
&&\omega^2(\ell,\lambda)=\omega_0^2\,(2\ell+3)(2\ell-1)(1-\lambda^{2\ell+1})\times\\
&&\left\{1-
\frac{\ell-\lambda^{2\ell+1}[(\ell+2)+(2\ell-1)(2\ell+3)-(2\ell+1)^2\lambda^2+(2\ell+3)\lambda^{2\ell+1}]
}{\quad\,
(2\ell-1)-\lambda^{2\ell+1}[(2\ell-1)(2\ell+3)-(2\ell+1)^2\lambda^2+(2\ell+3){\lambda}^{2\ell+1}]
} \right\}\\
\\
 \label{e2.15a}
 && \omega_0^2=\frac{c_t^2}{R^2}=\frac{\mu}{\rho}\quad \lambda=\frac{R_c}{R}=1-h\quad h=\frac{\Delta R}{R}.
\end{eqnarray}
In the limit of zero-size radius of the core, $\lambda=(R_c/R)\to
 0$, corresponding to torsional oscillations in the entire volume of
homogeneous elastic particle we regain the known result [6-8]:
\begin{eqnarray}
  \label{e2.21}
  \omega^2_t(\ell\geq 2,\lambda=0)=\omega^2_0[(2\ell+3)(\ell-1)].
 \end{eqnarray}
 One sees that in this limit the lowest overtone is again of quadrupole degree $\ell=2$.
 In the mean time, when the torsional vibrations are locked in the surface layer, the lowest overtone is of
 dipole degree and the frequency of $\ell=1$ torsional vibration is given by
 \begin{eqnarray}
  \label{e2.16}
 && \omega_t^2(\ell=1,\lambda)=\omega_0^2\,
 \frac{15\lambda^3(1-\lambda^3)}{(1-\lambda)^3(1+3\lambda+6\lambda^2+5\lambda^3)}.
\end{eqnarray}
 In Fig.3 we plot fractional frequencies of dipole both spheroidal
 $\omega_s(\ell=1)/\omega_0$ and  torsional $\omega_t(\ell=1)/\omega_0$ oscillations
 as functions of $h=\Delta R/R$,  the
 fractional thickness of the peripheral shell (see Fig.4), which is the measure of amount of mass that sets in vibrations.

 \begin{figure}
\centering{\includegraphics[width=10cm]{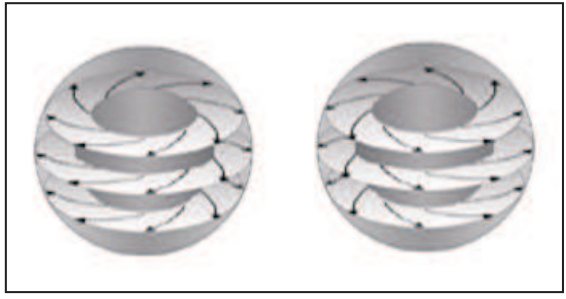}} \caption{
 Differentially-rotational displacements in dipole torsional soft mode of shear elastic oscillations of peripheral shell
 against core.
}
\end{figure}

 The smooth decreasing of frequency shows the more mass (volume) of nanoparticles sets in dipole
 vibrations of shell against core the less frequency of these vibrations.
 This figure
 clearly exhibits the Goldstone's soft-mode nature of dipole overtones of both spheroidal and
 torsional oscillations of the shell against core (Figs. 2 and 4).

\section{Summary}
   The two-component continuum model has been considered of a solid spherical nanoparticle, thought of as soft
   shell covering hard core, responding to optically induced perturbations by nodeless spheroidal and torsional
   shear elastic oscillations of peripheral finite-depth shell against hard immobile core.
   The practical usefulness of obtained
   Analytic equations for spectral formulae for the frequencies of the
   even-parity spheroidal (electric) spheroidal and the odd-parity (magnetic) torsional  modes
   trapped in the peripheral shell have been obtained in analytic form whose practical usefulness is that
   they provide a basis for constructive spectral analysis of
   resonance state of photoabsorption by ultrafine elastically deformable solid nanoparticles.
   Highlighted are
   the dipole overtones of nodeless elastic vibrations in the finite-depth peripheral layer as Goldstone's soft vibrational modes owing its existence to vibrations trapped in the peripheral shell, not in the entire volume of the particle.  The obtained spectral equations clearly show how the frequencies of optically induced elastic resonances
   depend upon particle material -- the shear modulus $\mu$, the density $\rho$ and
   on geometrical sizes of two-component nanoparticles -- depth of dynamical peripheral shell $\Delta R$ and the
   particle radius $R$.   Such information is indispensable to
   identification of experimentally observed
   picks of resonant photoabsorbtion by ultrafine micro and nanoparticles with eigenmodes of optically induced
   elastic oscillations (e.g.\cite{ADV-06,HHC-06}).
   Finally, we would like to point out that presented mathematical treatment of
   oscillatory behavior of a spherical mass of elastically deformable matter in the core-shell
   picture is quite general, so that obtained spectral formulae can found application to more wide class of physical
   objects of current basic and applied research.

\section{Acknowledgements}

  This work is partly supported by NSC of Taiwan, under grants NSC-
 97-2811-M-007-003 and NSC-
 96-2628-M-007-12-MY3.

\end{document}